# Cubic Rashba spin–orbit interaction of two-dimensional hole gas in strained-Ge/SiGe quantum well

Rai Moriya[1,a)], Kentarou Sawano[2], Yusuke Hoshi[1,2,\*)], Satoru Masubuchi[1,3], Yasuhiro Shiraki[2], Andreas Wild[4,5], Christian Neumann[5], Gerhard Abstreiter[4,6], Dominique Bougeard[5], Takaaki Koga[7], and Tomoki Machida[1,3,b)]

[1] *Institute of Industrial Science, University of Tokyo, 4-6-1 Komaba, Meguro-ku, Tokyo 153-8505, Japan*

[2] *Advanced Research Laboratories, Tokyo City University, 8-15-1 Todoroki, Setagaya-ku, Tokyo 158-0082, Japan*

[3] *Institute for Nano Quantum Information Electronics, University of Tokyo, 4-6-1 Komaba, Meguro-ku, Tokyo 153-8505, Japan*

[4] *Walter Schottky Institut and Physics Department, Technische Universität München, Am Coulombwall 4, 85748 Garching, Germany*

[5] *Institut für Experimentelle und Angewandte Physik, Universität Regensburg, Universitätsstraße 31, 93040 Regensburg, Germany*

[6] *Institute for Advanced Study, Technische Universität München, Lichtenbergstraße 2a, 85748 Garching, Germany*

[7] *Division of Electronics for Informatics, Graduate School of Information Science and Technology, Hokkaido University, Hokkaido 060-0814, Japan*

The spin–orbit interaction (SOI) of the two-dimensional hole gas in the inversion symmetric semiconductor Ge is studied in a strained-Ge/SiGe quantum well structure. We observed weak anti-localization (WAL) in the magnetoconductivity measurement, revealing that the WAL feature can be fully described by the *k*-cubic Rashba SOI theory. Furthermore, we demonstrated electric field control of the Rashba SOI. Our findings reveal that the heavy hole (HH) in strained-Ge is a purely cubic-Rashba-system, which is consistent with the spin angular momentum $m_j = \pm 3/2$ nature of the HH wave function.

a) E-mail: moriyar@iis.u-tokyo.ac.jp
b) E-mail: tmachida@iis.u-tokyo.ac.jp
*) Present address: Graduate School of Engineering, Nagoya University, Nagoya 464-8603, Japan

The spin–orbit interaction (SOI) in a two-dimensional system is a subject of considerable interest because the SOI induces spin splitting at zero magnetic field, which is important in both fundamental research and electronic device applications [1]. Recent developments of SOI-induced phenomena in solid state demonstrate many possibilities utilizing spin current and the emergence of new physics such as the spin interferometer [2,3], persistent spin helix [4,5], spin Hall effect [6,7,8], and quantum spin Hall effect [9,10]. Up to now, there have been two well-known SOIs existing in solids: the Dresselhaus SOI [11] due to bulk inversion asymmetry (BIA) in the crystal structure and the Rashba SOI [12,13] due to spatial inversion asymmetry (SIA).

In low-dimensional systems, the Rashba SOI becomes more important because it is stronger at the heterointerface and can be controlled by an external electric field. Many of the pioneering studies on SOI-induced phenomena mentioned above were performed in two-dimensional electron systems, where the Rashba SOI is described by the $k$-linear Rashba term. In the Hamiltonian, the $k$-linear Rashba term can be written as $H_{R1} = \alpha_1 E_z i(k_-\sigma_+ - k_+\sigma_-)$, where $\sigma_\pm = \frac{1}{2}(\sigma_x \pm i\sigma_y)$ denote combinations of Pauli spin matrices, $k_\pm = k_x \pm ik_y$, and $k_x$, $k_y$ are the components of the in-plane wave vector $\mathbf{k}_\parallel$. The effective magnetic field $\mathbf{\Omega}_1(\mathbf{k}_\parallel)$ acting on the transport carrier due to the $k$-linear Rashba term is illustrated in Fig. 1(a).

Recently, a higher-order contribution of the Rashba SOI, the so-called $k^3$ ($k$-cubic) Rashba SOI, has received more attention [14,15]. The Hamiltonian for the $k$-cubic Rashba SOI is expressed as $H_{R3} = \alpha_3 E_z i(k_-^3 \sigma_+ - k_+^3 \sigma_-)$ [16], and the effective magnetic field $\mathbf{\Omega}_3(\mathbf{k}_\parallel)$ in $k$-space is illustrated in Fig. 1(b) [15]. There is a significant difference in the effective field symmetry between the *k*-linear and the *k*-cubic Rashba SOI with one and three rotations in *k*-space, respectively. The $k^3$ symmetry of SOI is an interesting subject because it influences all of the SOI-induced phenomena instead of *k*-linear Rashba term. For example, in case of the spin Hall effect, the *k*-cubic Rashba term is predicted to give rise to a larger spin Hall conductivity [17,18,19].

Recently, the cubic-Rashba SOI has been reported on a two-dimensional hole gas (2DHG) in inversion asymmetric semiconductors InGaAs and GaAs [20,21], and a two-dimensional electron gas formed at a surface of inversion symmetric oxide $SrTiO_3$ [15]. However in the former case, the InGaAs and GaAs possess both BIA and SIA, thus they are always influenced by both Dresselhaus and Rashba SOI contributions [22]. Because the Dresselhaus SOI also has *k*-cubic dependence, the cubic-Rashba SOI is difficult to observe independently. In the latter case, a complex electric structure of the $SrTiO_3$ makes it difficult to compare with existing Rashba theory and theoretical development on the spin orbit interaction in perobskite-oxide is still in progress. Consequently, material systems with a simple band structure and pure cubic-Rashba SOI are in high demand. Here, we propose a purely cubic-Rashba SOI system: the single element semiconductor Ge. Because Ge does not have a BIA, the Dresselhaus SOI is essentially absent. Thus, the SOI in Ge is described only with Rashba term.

The valence band structure of Ge around the Γ point is schematically illustrated in Fig. 1(c). The valence band hole has a *p*-orbital wave function with orbital angular moment *l* = 1; therefore, it has total angular momentum $j = 3/2$ and $j = 1/2$. The $j = 3/2$ and $j = 1/2$ hole states are separated from each other due to the atomic SOI. While the separated $j = 1/2$ state is called the split-off (SO) state as indicated in the Fig 1(c). The $j = 3/2$ state further splits into the $m_j = \pm 3/2$ heavy hole (HH) and $m_j = \pm 1/2$ light hole (LH) states. Among the hole states, the wave function described by $m_j = \pm 1/2$ possesses $k$-linear Rashba SOI. The $k$-linear Rashba SOI is also dominant in most studied conduction band electron cases since conduction band is fully described by $m_j = \pm 1/2$ wave function. On the other hand, for $m_j = \pm 3/2$, Rashba SOI exhibits $k$-cubic dependence. Further, in the strained-Ge/SiGe quantum well (QW) structure, owing to compressive strain in the Ge well, the degeneracy of the HH and LH band at $k = 0$ can be lifted. The separation between HH and LH band can be as large as ~100 meV with increasing amounts of strain. Thus, the top of the valence band can be tuned to HH state. Recent advances in the Ge and SiGe epitaxial growth technology revealed that it is possible to fabricate high crystal quality strained-Ge/SiGe QW on Si substrates [23,24]. Thus, strained Ge/SiGe provides a good platform to study the cubic-Rashba SOI in simple single element semiconductors and will offer an interesting comparison with existing SOI theory.

In this Letter, we experimentally reveal cubic-Rashba SOI in 2DHG in Ge/SiGe QW. Comparing the weak localization (WL) and weak anti-localization (WAL) measurements with the analysis of the linear- and cubic-Rashba SOI terms, we found that spin transport in the 2DHG can be fully explained by the cubic-Rashba SOI. The SOI magnitude can be controlled with the gate voltage. These findings are all consistent with the $m_j = \pm 3/2$ characteristics of the HH band in strained-Ge QW, which obeys the cubic-Rashba Hamiltonian.

A strained-Ge/SiGe single QW is grown by using solid source molecular beam epitaxy. First, a fully relaxed $Si_{0.5}Ge_{0.5}$ buffer layer is grown on Si(001) substrate. On top of this layer, the following QW structure is grown: 10 $p$-$Si_{0.5}Ge_{0.5}$/10 $Si_{0.5}Ge_{0.5}$/20 Ge/30 $Si_{0.5}Ge_{0.5}$/3 Si (units in nm). Due to lattice mismatch (~2.1%) between the Ge and $Si_{0.5}Ge_{0.5}$ layer, the Ge QW layer is compressively strained. After patterning the wafer in the Hall bar structure, an $Al_2O_3$ dielectric layer with 75-nm thickness is deposited using atomic layer deposition, and then a Pd/Au gate electrode is deposited. The application of gate voltage enables us to tune the hole density of the 2DHG in the QW. The magnetotransport is measured in a variable temperature cryostat at the temperature range of 1.6 – 10 K. A magnetic field is applied perpendicular to the sample plane using a superconducting magnet.

First, the change in the sheet resistance $R_{sheet}$ and hole concentration $n_{hole}$ at various back gate voltage $V_G$ is shown in Fig. 2(a). When $V_G$ is decreased from +1.2 V to -0.4 V, $R_{sheet}$ ($n_{hole}$) monotonically decreases (increases). This change accompanies the variation of hole mobility from 2000 to 5200 cm$^2$/Vs. The change in the hole effective mass $m_{hole}$ with respect to $V_G$ is shown in Fig. 2(b). The effective mass of the holes are determined from the temperature dependence of the Shubnikov-de Haas oscillation. The effective mass increases with the hole concentration and this is due to the non-parabolicity of the valence band [25,26].

Next, we show the magnetoconductivity $\Delta\sigma(B) = \sigma(B)-\sigma(B=0)$ measured at various $V_G$ in Fig. 2(c). Under a gate sweep of $V_G$ = +1.0 V to -0.4 V, $\Delta\sigma(B)$ is significantly modulated. In the high-field region, the conductivity monotonically increases with the external magnetic field, which we attribute to the WL. Together with the WL feature, in the high hole density region ($V_G$ < +0.2 V) a conductance peak appears when the magnetic field is near zero. This conductance peak is due to the WAL. The cross-over points of $\Delta\sigma(B)$, which are indicated by arrows in Fig. 2(c), shift position from low magnetic field to high magnetic field with decreasing the $V_G$. The cross-over points are roughly proportional to the magnetic field for SOI, $B_{SO}$; the larger $B_{SO}$ produces a larger cross-over point field. Thus, our results suggest that the SOI can be tuned with gate voltage.

Observed WL and WAL data have been analyzed by using Iordanskii, Lyanda-Geller, Pikus (ILP) theory for magnetoconductivity which took into account both linear- and cubic-Rashba SOI [15,20,27]. The general formula of ILP theory is provided in the Supplemental Material [28]. We limited our fitting within diffusive transport regime where the magnetic field $B$ is within the range $B < B_{tr} = \hbar/2el_m^2$, where $B_{tr}$ denotes the transport field characterizing the elastic scattering time of the hole, $\hbar$ the Planck's constant, $e$ the electron charge, and $l_m$ the mean free path of the hole. In the ILP theory, we only consider the Rashba SOI since there is no structural inversion asymmetry in the Ge. Both $k$-linear $\Omega_1$ and $k$-cubic $\Omega_3$ effective magnetic fields in frequency units are incorporated [15]. Under these conditions, the fitting parameters to analyze experimental data are characteristic magnetic fields for the phase coherence $B_\phi = \hbar/4eD\tau_\phi$, for the $k$-linear spin-orbit coupling $B_{SO1} = \frac{\hbar}{4eD} 2|\Omega_1|^2 \tau_{tr}$, and for the $k$-cubic spin-orbit coupling $B_{SO3} = \frac{\hbar}{4eD} 2|\Omega_3|^2 \tau_{tr}$, where $D$ denotes the diffusion constant, $\tau_\phi$ the phase coherent time of the carrier, and $\tau_{tr}$ the elastic scattering time. By setting either $B_{SO3} = 0$ or $B_{SO1} = 0$ in the ILP model, the formulas with only the $k$-linear term or the $k$-cubic term are obtained [28]. In Fig. 3, we present the experimental data (open circles) and fitting results with only the $k$-linear term (blue dashed line) or only the $k$-cubic term (red solid line). We found that fitting with only the cubic-Rashba term revealed good agreement with the experimental results regardless of the $V_G$ value. Whereas, the linear-Rashba term is insufficient for explaining the larger field region of the WAL data. The low magnetic field region can be fitted with both the linear- and cubic- models because this region is determined by the $B_\phi$ term rather than $B_{SO}$. However, we observed significantly different behavior between the linear- and cubic-Rashba terms in the larger magnetic field region. The cubic-Rashba model explains the experimental data more precisely, thus we believe the cubic-Rashba model is relevant for the HH band in Ge.

Fig. 4(a) shows $B_{SO3}$ and $B_\phi$ obtained by the fitting with cubic-Rashba model. Both $B_{SO3}$ and $B_\phi$ systematically change with $V_G$. When $B_{SO3} > B_\phi$, WAL is observed, and WL is observed when $B_{SO3} < B_\phi$. The change of $B_{SO3}$ provides a direct evidence of electric field control of the cubic-Rashba SOI. Note that the change of $B_\phi$ with respect to the $V_G$ is due to the change of hole mobility, and this gives rise to a longer phase coherence time $\tau_\phi$ with increasing hole concentration. If $k$-dependent spin splitting due to SOI exists, the spin relaxation is described by the Dyakonov-Perel mechanism with $\tau_{SO} \propto 1/\Delta^2 \tau_{tr}$ [29], where $\Delta$ is the spin splitting energy and $\tau_{SO}$ denotes spin relaxation time. On the other hand, if spin-flip scattering is the dominant spin relaxation mechanism, the Elliot-Yafet mechanism with $\tau_{SO} \propto \tau_{tr}/(\Delta g)^2$ ($\Delta g$ is the shift of electron $g$ factor from that of free electrons) is relevant for describing our results. Therefore, the comparison between $\tau_{SO}$ and $\tau_{tr}$ provides additional evidence for Rashba-type SOI. The $\tau_{SO}$ and $\tau_{tr}$ terms can be calculated using the relations $B_{SO3} = \hbar/4eD\tau_{SO}$, $D = v_F^2 \tau_{tr}/2$, $v_F = \hbar k_F/m_{hole}$, and $k_F = \sqrt{2\pi n_{hole}}$, where $v_F$ is the Fermi velocity and $k_F$ is the Fermi wave vector. We plotted $\tau_{SO}$ and $\tau_{tr}$ with respect to the hole concentration as shown in Fig. 4(b). We found that $\tau_{tr}$ increases with increasing carrier concentration, whereas $\tau_{SO}$ decreases. This is consistent with the Dyakonov-Perel spin relaxation mechanism. Therefore, it provides further evidence that the Rashba SOI is a dominant spin–orbit contribution in the 2DHG.

The agreement between the experimental data and the *k*-cubic term only condition of the ILP theory reveals the existence of purely cubic-Rashba SOI in the 2DHG. We now evaluate the spin splitting energy from the relationship $\Delta = \hbar|\mathbf{\Omega}_3|$. The change of $\Delta$ with respect to hole concentration is shown in Fig. 4(c). We obtained $\Delta$ of 0.3 – 0.4 meV; we find larger spin splitting with increasing hole concentration. This spin splitting provides direct experimental evidence of the cubic-Rashba SOI induced splitting in strained-Ge. To analyze the splitting in more detail, we calculated the coefficient $\alpha_3 E_z$ for the cubic-Rashba SOI using the relation $\Delta = \hbar|\mathbf{\Omega}_3| = \alpha_3 E_z k^3$, as shown in Fig. 4(d). The coefficient $\alpha_3 E_z$ decreases with increasing hole concentration. We believe that this is due to the change of gate-induced electric field $E_z$ in the QW rather than change of $\alpha_3$.

Here, the Rashba coefficient $\alpha_3$ for the first HH sub band is expressed as follows:

$$\alpha_3 = \frac{-3e\hbar^4 \gamma_3^2}{2m_0^2 (E_{HH} - E_{LH})^2} \qquad (0)$$

where $\gamma_3$ denotes the Luttinger parameter, $E_{HH}$ and $E_{LH}$ are the energies of the lowest HH and LH sub-band, respectively. Eq. (1) is derived from the Luttinger Hamiltonian using the partitioning method presented by Ohkawa and Uemura [13]. Eq. (1) suggests that $\alpha_3$ is inversely proportional to the square of the separation between the HH and LH energy. Qualitatively similar expression with eq. (1) is also reported by other groups [14,20]. In a single heterostructure, the HH-LH separation changes with the external electric field, thus demonstrating gate-control of $\alpha_3$ [30]. However, in the QW, the change in HH and LH separation with external electric field should be small. Furthermore, because the HH-LH separation in the Ge/Si$_{0.5}$Ge$_{0.5}$ QW is initially large due to strain, the gate-induced change of HH-LH separation is one order smaller than its total HH-LH separation energy; therefore, we think $\alpha_3$ is almost constant over the studied hole concentration range. For these reasons, we believe that the results shown in Fig. 4(d) reflect the change of effective electric field in the QW under the application of gate voltage. Because the gate-induced electric field in the QW opposes the initial electric field due to modulation doping on the bottom side of the QW, the increase of the hole concentration in the QW decreases its internal electric field. Thus, $\alpha_3 E_z$ decreases with increasing hole concentration.

Eq. (1) suggests another interesting aspect of the cubic-Rashba coefficient. The Rashba coefficient can be strongly tuned with the HH-LH separation. We estimated $\alpha_3$ using Eq. (1) with $\gamma_3 = 5.69$ for Ge [1]. The HH-LH splitting for compressively strained-Ge/Si$_{0.5}$Ge$_{0.5}$ is ~110 meV [31]. We obtained $\alpha_3 = 2.26 \times 10^5$ eÅ$^4$. The coefficient $\alpha_3$ is about one order smaller than the 2DHG in the GaAs/AlGaAs single heterostructure [21], which is a nearly lattice matched system. The smaller $\alpha_3$ value is due to the large HH-LH separation in Ge/Si$_{0.5}$Ge$_{0.5}$ compared to that of the GaAs/AlGaAs structure; large HH-LH separation decreases the coefficient $\alpha_3$. In the strained-Ge/SiGe QW, the HH-LH separation can be tuned by the strain due to the SiGe layer; in other words, the composition $x$ of Si$_{1-x}$Ge$_x$ buffer layer enables a wide range of tuning options for cubic-Rashba SOI. Note that Ge has larger $\gamma_3$ value than Si ($\gamma_3 = 1.45$) [1]; thus we expect larger SOI for 2DHG in Ge than that of Si. Therefore 2DHG in strained-Ge is promising material for utilizing cubic-Rashba SOI in low dimensional system.

**Acknowledgements**

We appreciate the valuable discussions with H. Nakamura. This work was partly supported by Grants-in-Aid for Scientific Research (Grant Numbers 24684024, 24340065, 25600079, and 26286044) from the Japan Society for the Promotion of Science (JSPS); the Project for Developing Innovation Systems of the Ministry of Education, Culture, Sports and Technology; the Deutsche Forschungsgemeinschaft via SFB 631; Deutsche Forschungsgemeinschaft via SFB 631; the Nanosystems Initiative Munich; and the Hitachi Metals·Materials Science Foundation.

Figure captions

Fig. 1

(Color online) (a,b) The effective magnetic field direction with respect to the in-plane wave vector $k$ for (a) the linear-Rashba term, $\mathbf{\Omega}_1$ and (b) the cubic-Rashba term, $\mathbf{\Omega}_3$. (c) Schematic illustration of the valence band structure for Ge. The HH-, LH-, and SO-band are indicated.

Fig. 2

(Color online) (a,b) The $V_G$ dependence of (a) sheet resistance $R_{sheet}$ and hole concentration $n_{hole}$ and (b) the hole effective mass $m_{hole}$. (c) The change of magnetoconductance $\Delta\sigma(B)$ measured at 1.6 K at various $V_G$.

Fig. 3

(Color online) The $V_G$ dependence of magnetoconductance $\Delta\sigma(B)$ shown with the fitting results from the ILP model. Black open circles indicate experimental data, and the red solid line and blue dashed lines show calculations using the ILP model for the cubic- and linear-Rashba terms, respectively.

Fig. 4

(Color online) (a) The $V_G$ dependence of the characteristic magnetic fields for the $k$-cubic spin-orbit coupling $B_{SO3}$ and the phase coherence $B_\phi$. (b) Hole concentration dependence of the spin relaxation time $\tau_{SO}$ and momentum scattering time $\tau_{tr}$. (c) Spin-splitting energy Δ versus the hole concentration. (d) The change of Rashba spin–orbit coefficient $\alpha_3 E_z$ with respect to the hole concentration.

Figure 1

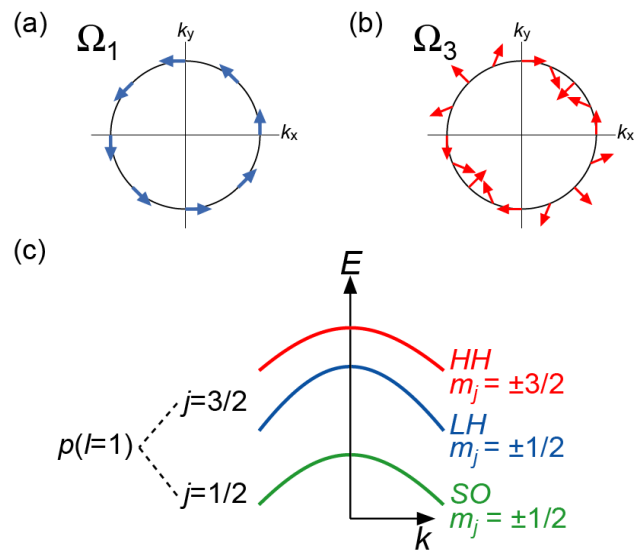

Figure 2

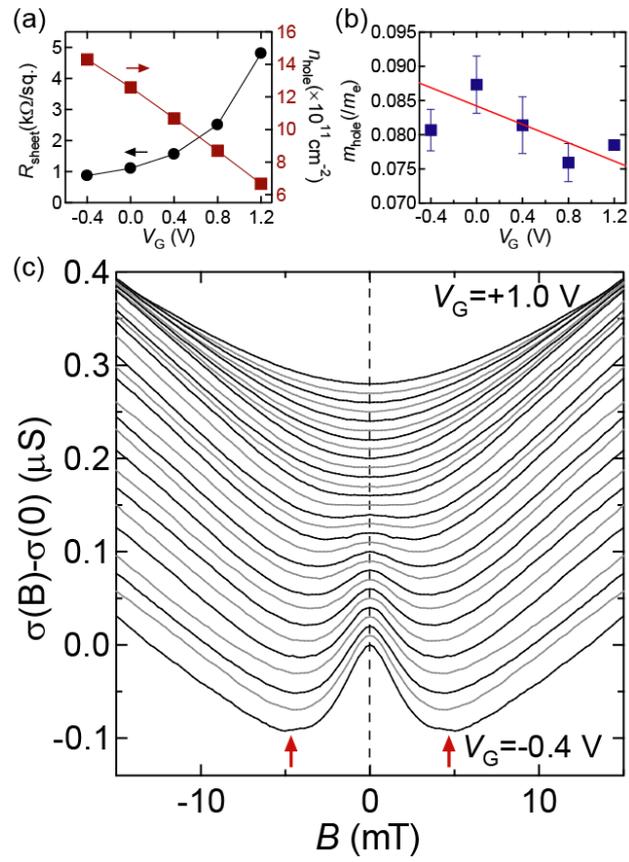

Figure 3

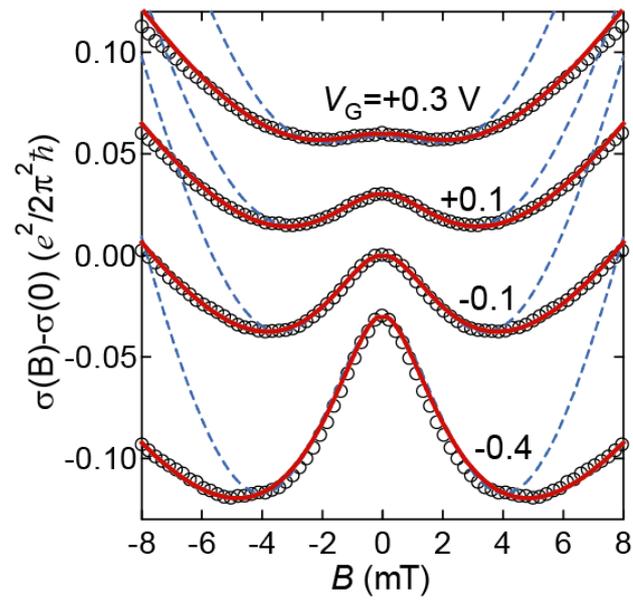

Figure 4

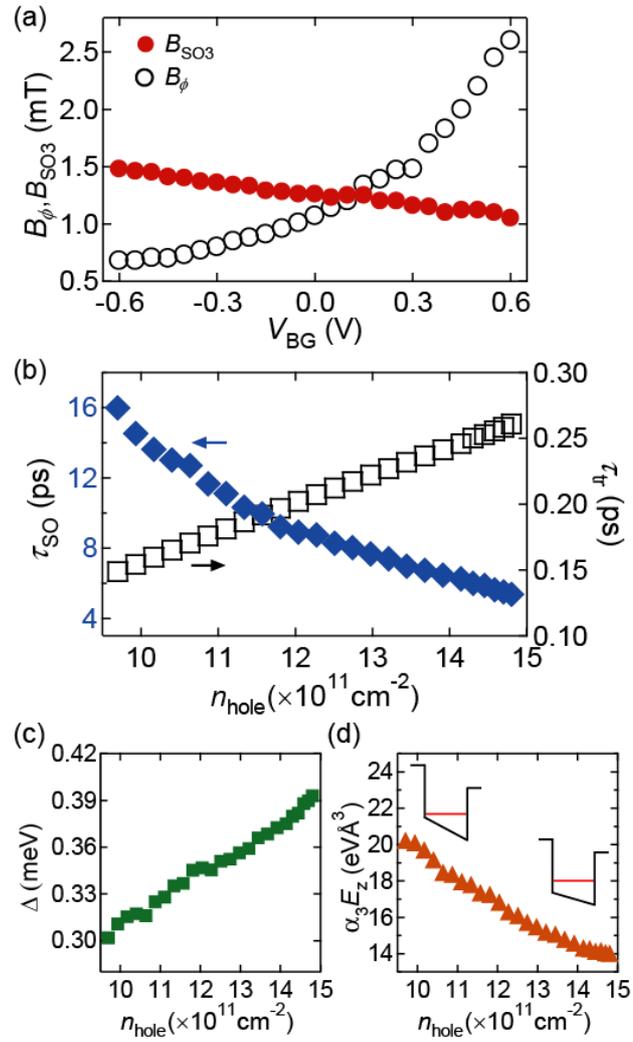